\documentclass[letterpaper]{article} 
\usepackage[draft]{bbx/aaai2026}  
\usepackage{times}  
\usepackage{helvet}  
\usepackage{courier}  
\usepackage[hyphens]{url}  
\usepackage{graphicx} 
\usepackage{natbib}  
\usepackage{caption} 
\usepackage{algorithm}
\usepackage{algorithmic}

\usepackage{subcaption}
\newcommand{\oursys}{BodyWave}
\usepackage{amsmath}

\usepackage{newfloat}
\usepackage{listings}
\DeclareCaptionStyle{ruled}{labelfont=normalfont,labelsep=colon,strut=off} 
\floatstyle{ruled}
\newfloat{listing}{tb}{lst}{}
\floatname{listing}{Listing}
\title{BodyWave: Egocentric Body Tracking using mmWave Radars on an MR Headset}
\author{
    Yin Li\textsuperscript{\rm 1, 2},
    Sean Korphi\textsuperscript{\rm 1}, 
    Sam Shiu\textsuperscript{\rm 1}, 
    Yasuo Morimoto\textsuperscript{\rm 1} \\
    Jiang Zhu\textsuperscript{\rm 1}, 
    Rajalakshimi Nandakumar\textsuperscript{\rm 2}, 
    John Ho\textsuperscript{\rm 1} \\
    \textsuperscript{\rm 1} Meta Reality Lab  
    \textsuperscript{\rm 2} Cornell University
}

\begin{document}

\maketitle

\begin{abstract}
Egocentric body tracking, also known as inside-out body tracking (IOBT), is an essential technology for applications like gesture control and codec avatar in mixed reality (MR), including augmented reality (AR) and virtual reality (VR). However, it is more challenging than exocentric body tracking due to the limited view angles of camera-based solutions, which provide only sparse and self-occluded input from head-mounted cameras, especially for lower-body parts.
To address these challenges, we propose, \oursys, an IOBT system based on millimeter-wave (mmWave) radar, which can detect non-line-of-sight. It offers low SWAP+C (size, weight, and power consumption), robustness to environmental and user factors, and enhanced privacy over camera-based solutions.
Our prototype, modeled after the Meta Quest 3 form factor, places radars just 4cm away from the face, which significantly advances the practicality of radar-based IOBT.
We tackle the sparsity issue of mmWave radar by processing the raw signal into high-resolution range profiles to predict fine-grained 3D coordinates of body keypoints. In a user study with 14 participants and around 500,000 frames of collected data, we achieved a mean per-joint position error (MPJPE) of 9.85 cm on unseen users, 4.94 cm with a few minutes of user calibration, and 3.86 cm in a fully-adapted user-dependent setting. This is comparable to state-of-the-art camera-based IOBT systems, introducing a robust and privacy-preserving alternative for MR applications.
\end{abstract}

\section{Introduction}

\begin{figure*}[!ht]
    \begin{minipage}{0.645\textwidth}
        \centering
        \includegraphics[width=0.9\linewidth]{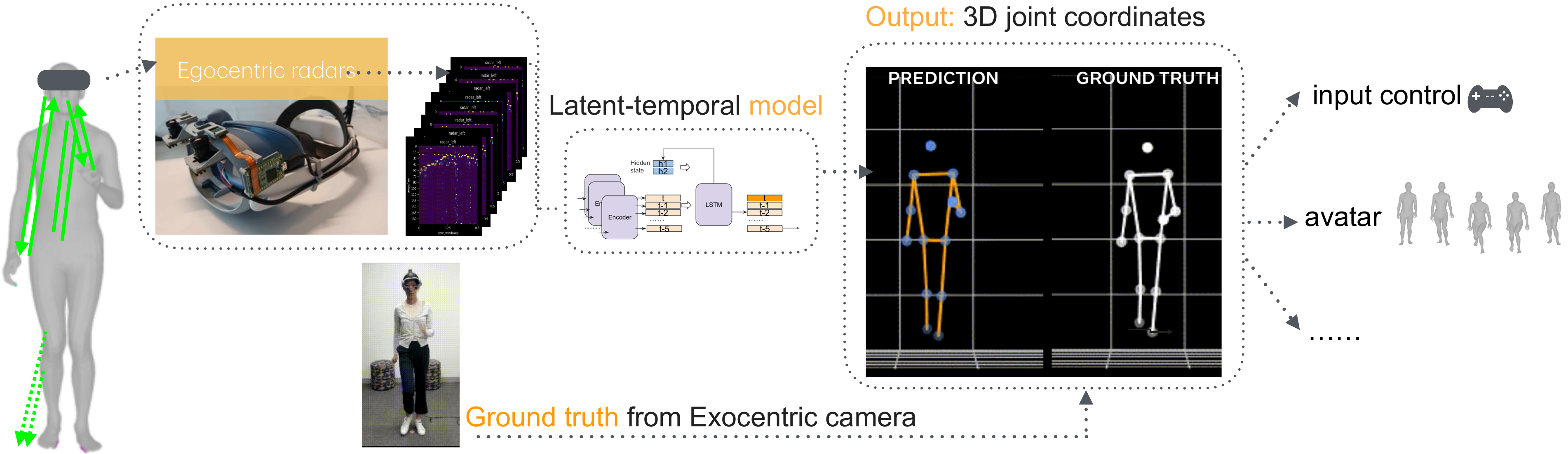}
        \vspace{-8pt}
        \caption{System overview.}
        \label{fig:teaser}
    \end{minipage}
    \hfill
    \begin{minipage}{0.335\textwidth}
       \centering
        \includegraphics[width=.99\linewidth]{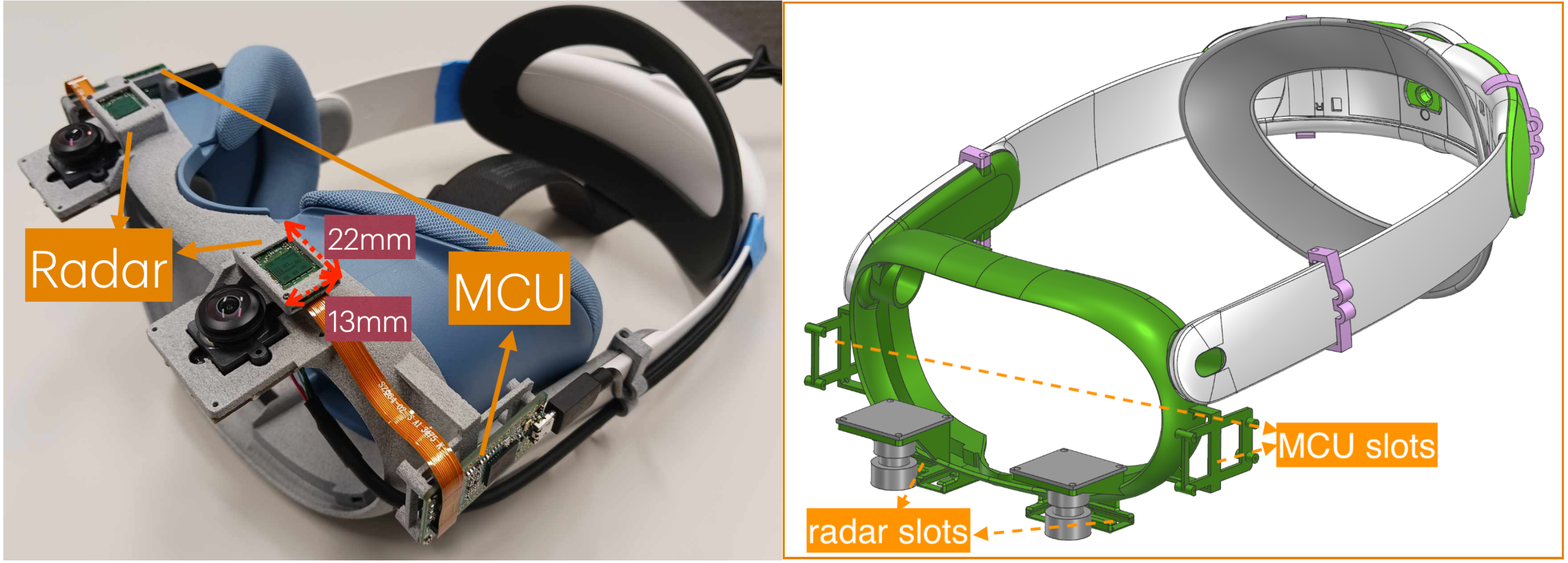}
        \vspace{-7pt}
        \caption{Our headset prototype.}
        \vspace{-7pt}
        \label{fig:prototype}
    \end{minipage}
    \vspace{-15pt}
\end{figure*}

IOBT captures body movements from a first-person perspective, using wearables like smart glasses and MR headsets. This approach contrasts with exocentric body tracking, which relies on external sensors or cameras.
IOBT holds significant potential for applications in virtual communication, such as codec avatar, a photorealistic digital representation of individuals. 
Since about 70\% of communication is non-verbal, accurately capturing gestures is crucial to conveying emotion and intent in virtual interaction.
Other potential applications of IOBT include gesture control for MR gaming and healthcare, such as physical therapy and rehabilitation.

Various sensing modalities have been explored for IOBT, including wearable motion capture systems utilizing IMUs or LED markers, as well as head-mounted infrared (IR) or fisheye cameras. Among these, head-mounted cameras are particularly promising, given the growing popularity of smart glasses and MR headsets such as Meta Quest, Apple Vision, and PICO.
However, current commercial devices support only coarse-grained IOBT, such as hand gesture classification (Apple Vision) and 3-point body tracking (Meta Quest). This limitation arises from several challenges:
1) Limited field of view, even with fisheye lenses, hampers full-body capture—especially the lower body;
2) Self-occlusion, as limbs block other body parts during movement;
3) Varying lighting, which degrades image quality and hinders tracking accuracy.
Camera-based IOBT is currently a rising field, but it remains less developed than exocentric body tracking, largely due to the lack of realistic and practical datasets. They are either synthetic data~\cite{akada2022unrealego, xu2019mo}, or missing accurate egocentric body poses~\cite{grauman2022ego4d, khirodkar2023ego, wang2022estimating}, or built upon head-mounted cameras that are impractically far away from the body~\cite{rhodin2016egocap, tome2020selfpose, tome2019xr, xu2019mo}. Moreover, the shift toward pancake lenses in commercial MR headsets—aimed at reducing device thickness—further narrows the field of view, hindering fine-grained tracking.

Therefore, we aim at using wireless perception for IOBT as a superior solution, overcoming angle and occlusion limitations of camera-based systems. Wireless perception typically leverages modalities such as radio frequency (RF) and ultrasound, with mmWave being particularly promising for IOBT due to its compact size, suitable detection range, and high resolution.
Specifically, 1) it enables non-line-of-sight (NLOS) sensing, allowing detection of occluded body parts, such as the lower body.
2) mmWave radar systems have a lower SWAP+C (size, weight, power, and cost), which is suitable for headsets or glasses, where space and power are limited.
3) it is more robust to environmental and user factors than cameras, such as conditions of poor lighting, smoke, or fog.
4) it poses fewer privacy concerns, as it does not capture visual images of people or surroundings.


Early work has demonstrated the potential of mmWave technology for IOBT~\cite{mmEgo}, but it requires fusion with IMUs to detect head rotation and are impractical for commercial MR devices due to the radar's large size and long distance from the face. Specifically, the radar must be placed unrealistically far away from the face (20cm+) to maximize the view of the body, mounted to a front plate attached to a helmet. However, recent advancements in optics, such as pancake lenses used in Meta Quest and Apple Vision Pro, reduce headset thickness to around below 9 cm.
Furthermore, its studies were limited to only 3 test users and requires at least 2 meters of surrounding clearance, significantly restricting real-world deployment.


In this paper, we propose \oursys, the first mmWave-based IOBT system integrated into an MR headset. Our design addresses key limitations of prior work and offers several advantages:
1) Compact and close to the face: the prototype headset is in the same form factor of Meta Quest 3, equipped with two downward-facing Infineon BGT60ATR24C mmWave radars, at the same height as the nose bridge (4-5cm away from the cheek). This makes it practical for commercial MR headsets or even smart glasses.
2) Fine-grained tracking: as show in Fig.~\ref{fig:teaser}, \oursys{} leverages a custom high-resolution radar range profile and a latent-temporal model to predict 13 full-body key points. The supervised ground truth is from an exocentric RGB camera placed in the front of the user, which is only required for training not testing.
To evaluate its performance, we conducted extensive experiments with 14 participants performing 19 selected gestures, involving camera-challenging leg gestures, hand-body contact, and head rotations.
\oursys{} achieves an MPJPE of 4.94 cm on users with a few minutes of adaptation/calibration data, and 9.85 cm for brand new participants with no seen data in training. The fully-adapted user-dependent case could further decrease to 3.86 cm.
These results demonstrate the effectiveness of the proposed mmWave IOBT, even in challenging scenarios, making it a promising complementary modality beyond cameras, especially for highly-occluded lower-body tracking. 

In the following, we first detail our methodology, study design, and benchmarking results across various users, rooms, and other factors. We also demonstrate a downstream application by applying our predictions to a body codec avatar, followed by discussions and future work.

\section{Method}
\subsection{Prototype design}
\label{sec:prototype_design}

    

\begin{figure}
    \centering
    \includegraphics[width=0.55\linewidth]{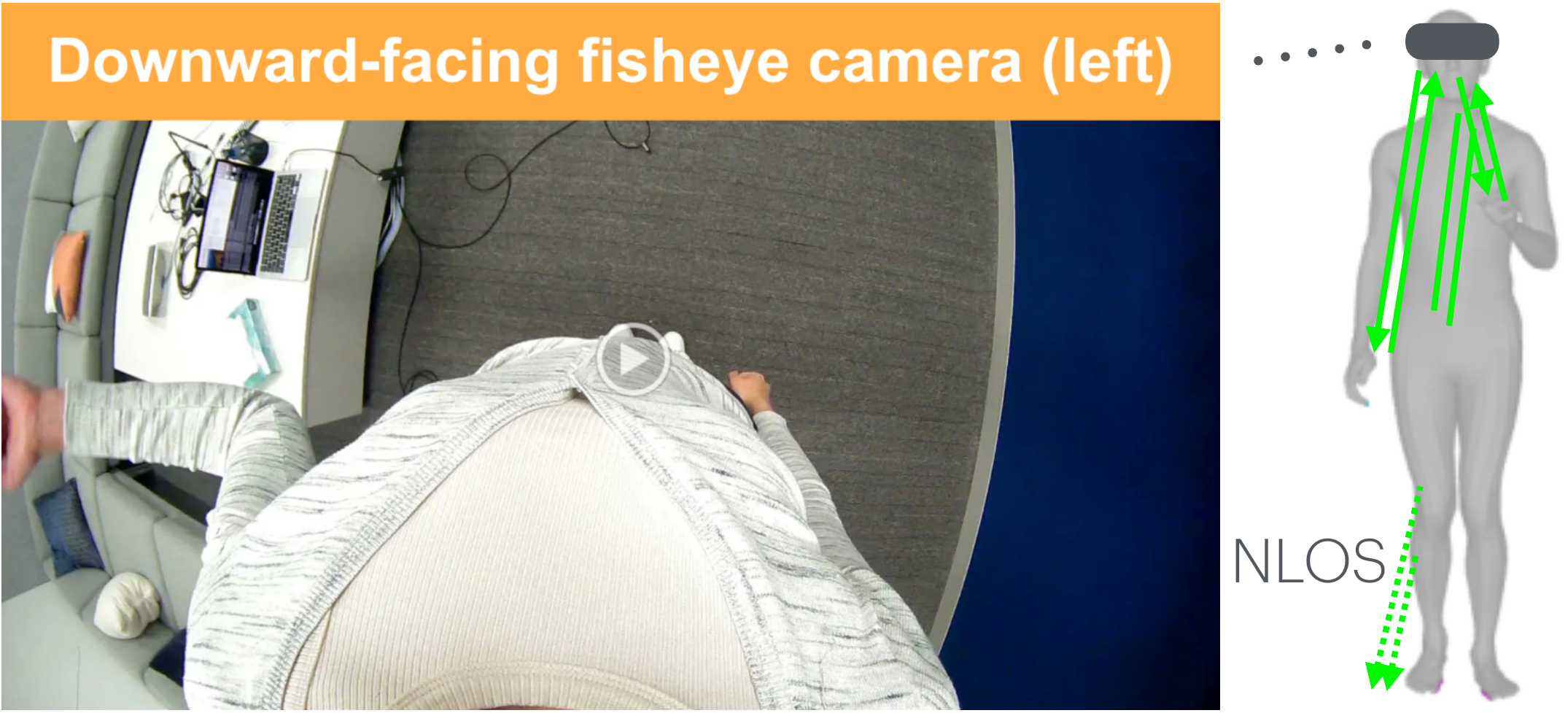}
    \vspace{-7pt}
    \caption{Camera's limited field-of-view and none-line-of-sight occlusions.}
    \vspace{-10pt}
    \label{fig:fisheye}
\end{figure}

As in Fig.~\ref{fig:prototype}, the prototype headset is in the same form factor of Meta Quest 3, equipped with two additional downward-facing mmWave radars on the bottom side.
\textbf{Radar: }
The radars are Infineon XENSIV BGT60ATR24C 60 GHz~\cite{infineonRadar}; each supports 4 GHz bandwidth with 2 TX / 4 RX channels, enabling ultra-wide bandwidth FMCW housed in a compact 22mm X 13mm PCB.
The radars are mounted at the left and right bottom edge of the headset. Both face downward, aligned with the nose bridge and just 4cm away from the cheek. This makes them more suitable for integration into commercial MR headsets or even smart glasses like Meta Orion ~\cite{orion2024}, in line with whose vision of replacing MR headsets.
In contrast, cameras suffers from limitations at similar deployment positions, due to field of view (FoV) and occlusion of lower body.
Fig.~\ref{fig:fisheye} is a sample view positioned next to our radars. Even a fisheye camera with a 200-degree horizontal FoV located 6.5cm from the cheek~\cite{CaliCam} fails to capture the self-occluded lower body and extended hand. Notably, this FoV is already the widest available, optimized by SLAM algorithms.
(Fisheye camera is not used for any training or testing in this paper, but purely included for comparison or possibly future work on sensor fusion.)
\textbf{Headset: }
It is composed of a Meta Quest 3 head strap, a facial interface, and a 3D-printed frame (HP MJF Nylon) to house the sensors, all maintaining the form factor of the Meta Quest 3.
Each radar has a dedicated microcontroller unit (MCU) and is USB-connected to a laptop. In the future, the MCU can be replaced with the existing compute on headsets.
\textbf{Ground truth camera: }
For ground truth only, an exocentric/outside-in RGB camera ~\cite{logitechCamera} captures 1920×1080@30FPS video from the front of the user. 
Then, we use MediaPipe body API~\cite{mediapipe} to extract the ground truth body keypoints. No camera is needed during inference.

\subsection{Understanding our high-resolution signal design}
Typical mmWave systems utilize Frequency Modulated Continuous Wave (FMCW) modulation techniques to estimate range, velocity, and angle. Many radar software development kits (SDK) further derive 3D point clouds, which can be sparse and lossy.
In contrast, we process raw signal into high-resolution range profiles, basically multi-channel 2D feature maps, as input to machine learning model.
In the following, we detail our signal processing design and custom enhancements to maximize detection resolution, followed by visualizations to validate and interpret the effectiveness of range profile as our core feature representation.

\begin{figure}
    \centering
    \includegraphics[width=0.8\linewidth]{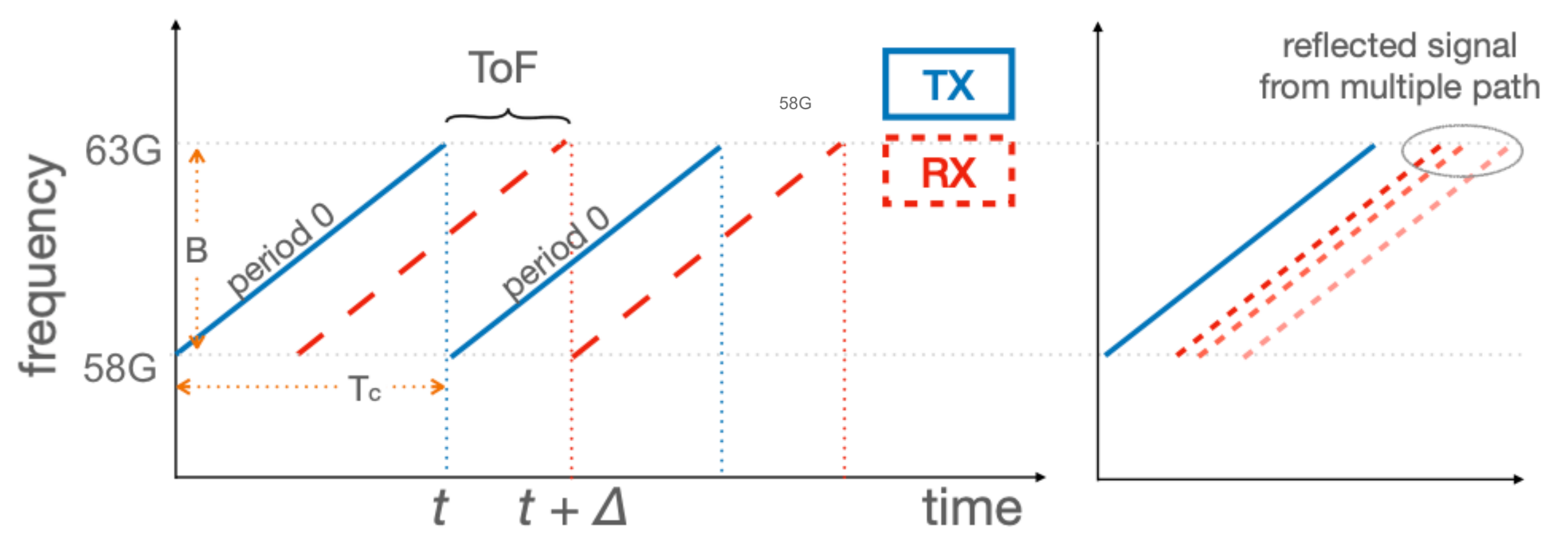}
    \vspace{-7pt}
    \caption{FMCW signal: each received signal is a time-delay version of the transmitted signal shifted by a different amount of time proportional to the distance.}
    \label{fig:fmcw}
    \vspace{-10pt}
\end{figure}

\textbf{FMCW signal modulation:}
An FMCW radar continuously transmits a frequency-modulated signal and receives reflections from the surroundings. In essence, each received signal is a time-delayed version of the transmitted signal, with the delay proportional to the distance of the object. 
To elaborate, as in Fig.~\ref{fig:fmcw}, FMCW modulation involves linearly varying the frequency over time, resembling repeated chirps. This can be expressed as: $A\sin(2\pi( f_{0}t + \frac{f_1-f_0}{2T}t^2))$, where $f_0$ and $f_1$ are the starting and ending frequency of the chirp; $T$ is the time duration of one chirp.
On the receiver (RX) side, intermediate frequency (IF) is generated by mixing the received signal with transmitted (TX) signal. After amplification, filtering, and digitization, the IF signal encodes the time of flight, allowing computation of round-trip distance using the known signal propagation speed.

\textbf{Signal processing and optimizations:} 
Next, raw signals can be processed into 2D range profiles, velocity, angle of arrival, or further combined into 3D point clouds. We prefer range profiles, when determining the optimal signal processing pipeline as input into downstream deep learning algorithms for body tracking; even though point clouds are the default in most radar SDKs. Because range profiles offer significant advantages:
1) fine-grained information of multi-targets, unlike point clouds which are lossy and sparse due to the signal filtering through peak detection.
2) additional temporal information per gesture within a single spectrogram.
3) less computational consumption, making it efficient for real-time tracking applications. 
We also employ two custom optimizations to enhance its fidelity:
1) Successive subtraction, that suppress noise from static environmental reflections. This involves subtracting each data column from the preceding one, assuming that user can never appear stationary; because there will always be minor body jitters, and our advanced detection resolution is 3 cm.
2) Frame unwarping, which optimizes temporal resolution for Fast Fourier Transform (FFT), by processing each chirp independently rather than aggregating chirps per frame. This maximizes chirp rate, thereby enhancing the temporal granularity of data.

\textbf{Configuration and resolution:} 
Our radar is configured with an analog-to-digital converter (ADC) sample rate of 1M Hz and a frame rate of 30 Hz, matching that of the ground truth camera. Each frame comprises 4 chirps, each with 128 samples sweeping from 58 GHz to 63 GHz, yielding a substantial frequency bandwidth of 5 GHz. The chirp repetition time is 5e-4 seconds, in compliance with sample rate and frame repetition.
All 4 transmitters and 2 receivers are activated (bit masks: 15 and 3), resulting in a total of 8 virtual channels.
The band-pass filter is between 80KHz and 500KHz (half the sample rate) for anti-aliasing.
The TX power level is set to the maximum unit value, 31 (not in decibels (dB)). The IF gain is 33 dB, safely within Federal Communications Commission (FCC) limits.
Therefore, the detection resolution, i.e. the minimal displacement detectable by our radar, is calculated as $\frac{c}{2B} = 3e8/(2*5e9) =0.03m$ where $B$ is bandwidth and $c$ is the speed of radio signal.
Accordingly, the maximum detection range is $128*0.03m/2 = 1.92m$ that aligns well with IOBT use case.

\begin{figure}[H]
    \centering
    \vspace{-7pt}
    \includegraphics[width=0.7\linewidth]{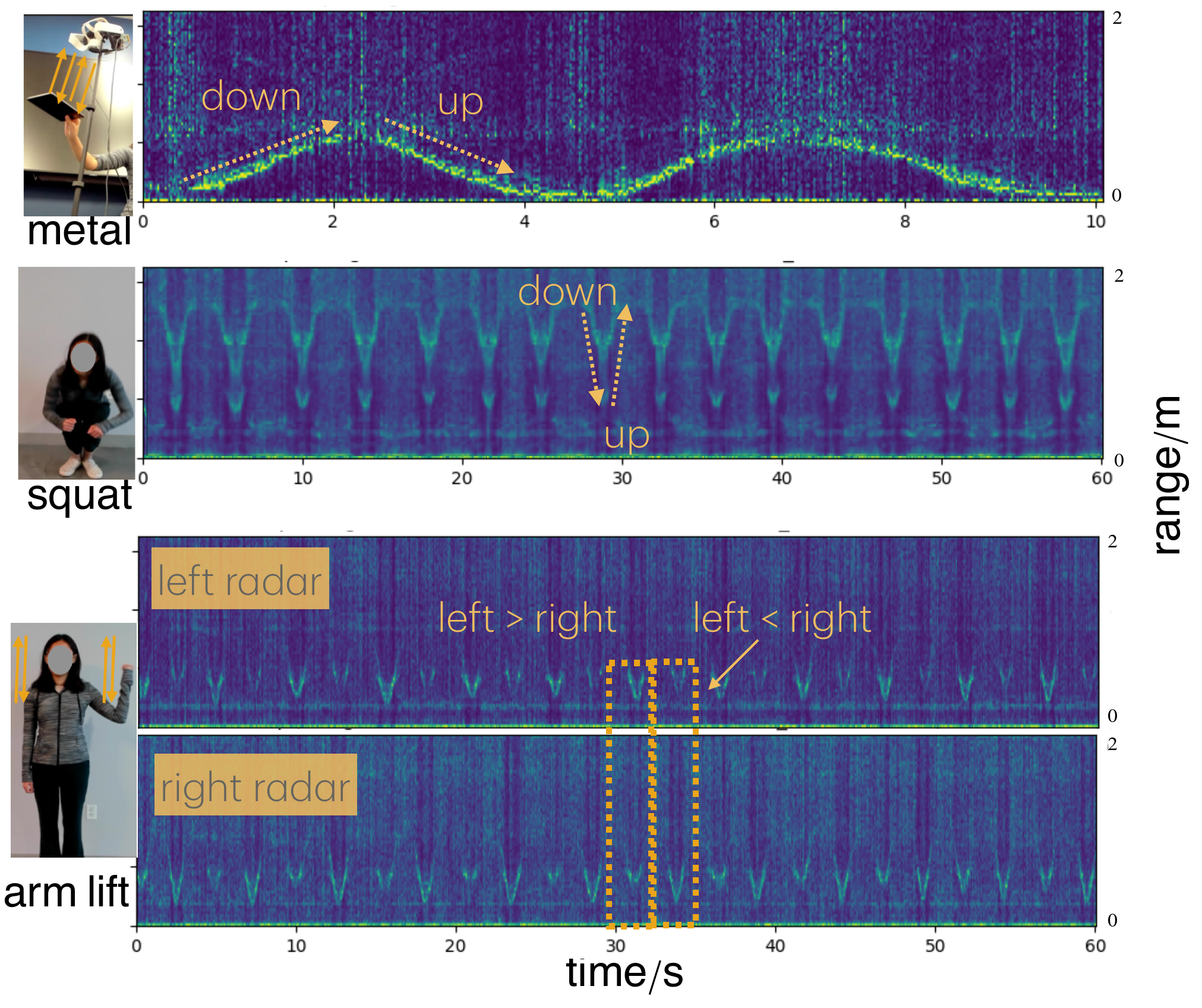}
    \vspace{-7pt}
    \caption{Interpreting range profiles via controlled motions.}
    \vspace{-7pt}
    \label{fig:spectrograms}
\end{figure}

\textbf{Interpreting the range profile:}
In essence, range profile is a 2D depth spectrogram; multiple channels encompass multi-view 3D information. It reflects the presence of surfaces at varying distances over time, thereby revealing patterns associated with different gestures.
To understand the patterns a deep learning model can perceive from spectrograms, we conduct experiments with 3 controlled motions, and then visualize and interpret the spectrograms, as in Fig.~\ref{fig:spectrograms}(Humans are blocked per double-blind requirement):
1) Flat metal (an iPad) moving back and forth in 1D is straightforward for radar, due to its strong reflections. As depicted in the first row, distinct highlights correspond to the metal’s motion; the highlight rises along the time axis as the metal moves down away from the radar and decreases when moving closer. Two crests appear as the object moves back and forth twice, at around 1 meter away.
2) When squatting repeatedly with \oursys{} mounted on the head, the second row illustrates similar crests of highlights but over a broader range (0-2 meters). The spectrogram appears noisier due to the complex multi-path reflections from various body parts, yet consistent motion-induced crests remain visible.
3) Arm lifting is shown in the last two figures, each representing one channel from the left or right radar. There are similar crests but within a closer range, with asymmetries reflecting arm movements: the left radar shows a larger crest when the right arm is lifted, and vice versa.
In summary, this analysis helped validate the spectrogram’s relevance to body gestures and informed the design of our deep learning model.

\subsection{Multi-view latent-temporal model}
\textbf{}



\begin{figure}[H]
    \centering
    \vspace{-7pt}
    \includegraphics[width=0.8\linewidth]{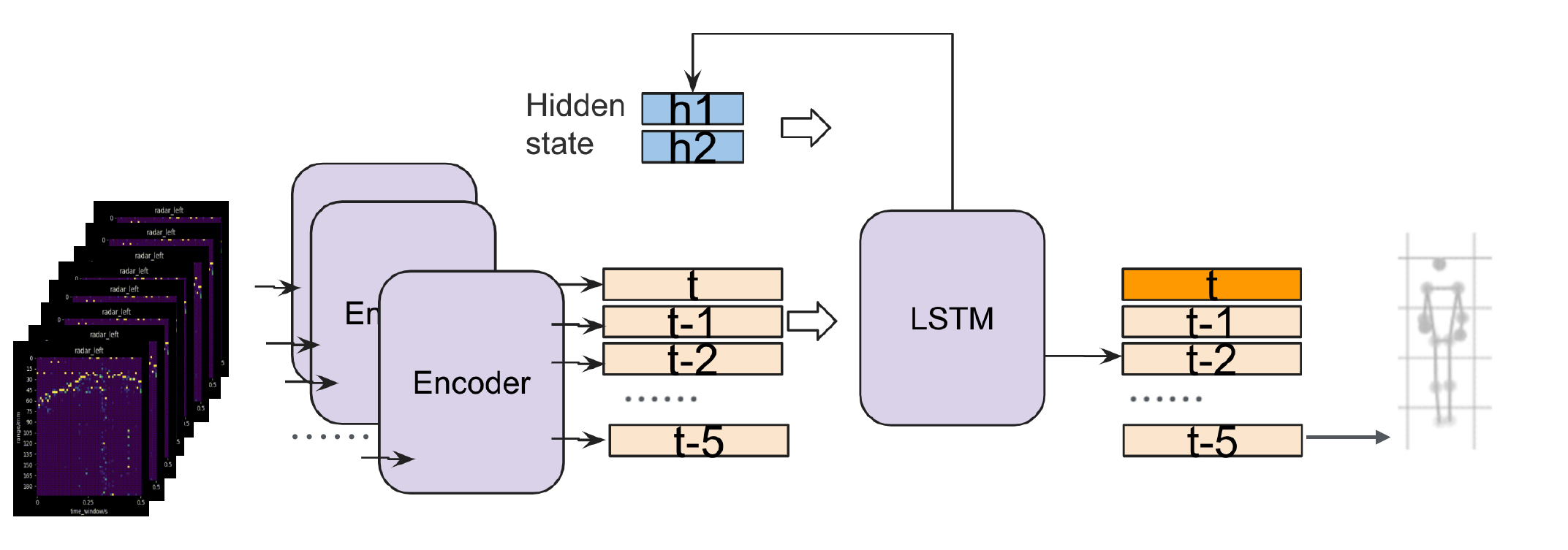}
    \vspace{-10pt}
    \caption{Architecture of the latent-temporal model.}
    \vspace{-10pt}
    \label{fig:model_arch}
\end{figure}

To design a deep learning model tailored for radar range profiles and IOBT, we evaluated three architectures: 1) a latent-temporal model, 2) an adapted cross-view transformer, 3) a vanilla convolutional neural network (CNN) model. Under consistent experimental conditions (detailed in the following section, per user-adaptive testing), the latent-temporal model outperformed the others and is selected as our final design, with errors of 4.94, 7.45, 14.72cm separately.
As demonstrated in Fig.~\ref{fig:model_arch}, the input layer has 8 channels, each with a 60x65 feature map. Each channel is segmented into 6 windows and feed into the same two CNN layers with weight sharing. Each CNN layer comprises BatchNorm, ReLU activation, and max pooling. 
In sequence, all windows are concatenated to feed into an LSTM model following by linear layers to match the output shape. This architecture allows the LSTM to capture latent-temporal features and make smooth and accurate predictions about the body's movements over time. The output of the transformer is a set of 13 3D keypoints.
The hyperparameters include a batch size of 512, 50 epochs, an initial learning rate of 0.01, plus a ReduceLROnPlateau learning rate scheduler (factor 0.5, patience 5 epochs, minimum LR 1e-5).
Overall, the model captures spatial and temporal features from multiple channels of radar data over time, enabling smooth, accurate tracking with minimal jitter.

\begin{figure}[H]
    \centering
    \vspace{-10pt}
    \includegraphics[width=0.75\linewidth]{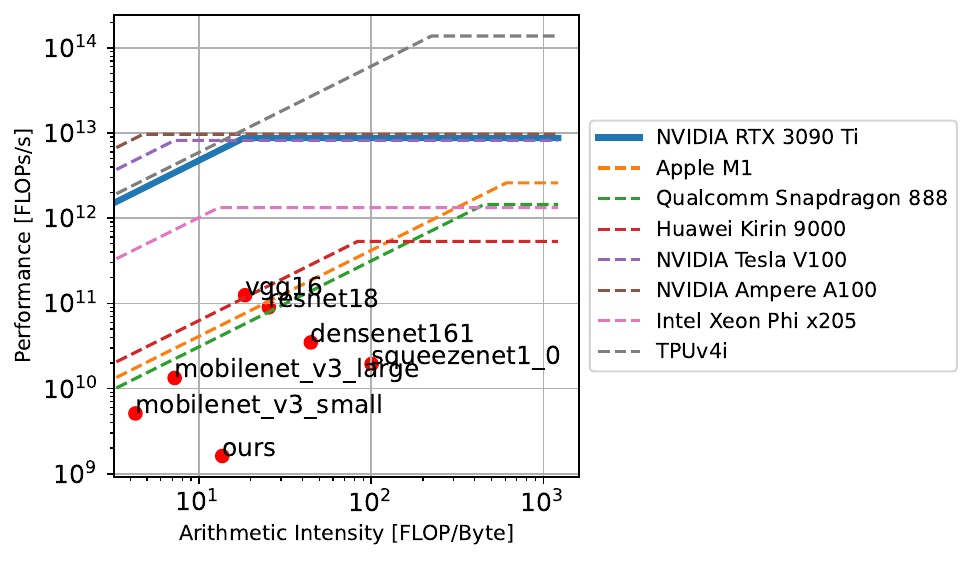}
    \vspace{-10pt}
    \caption{Roofline plot for model efficiency.}
    \vspace{-7pt}
    \label{fig:roofline}
\end{figure}

\subsection{Computation efficiency}
\textbf{}

\oursys{} is designed for real-time operation by processing buffered radar streams segmented into small windows. In this study, we use a 0.5-second window, balancing motion capture granularity and computational efficiency.
The model has a compact size of 209.5KB, with a forward/backward pass of 3.98MB and an input size of 1.21MB, totaling around 5.19MB. So, inference can be well within the memory capacity of commercial mobile devices. Although trained on an RTX 4080 GPU, the model can perform inference at approximately 325 frames per second (FPS) on mobile devices, equating to about 3.07 milliseconds per frame. Given the 0.5-second buffer interval, there is ample time for inference on each frame, allowing real-time performance without noticeable delay.
While further improvements via model compression and inference acceleration are possible, they are beyond the scope of this work.
We also present a roofline plot in Fig.~\ref{fig:roofline}, comparing our model with other popular models on an RTX 3090 GPU. Our model is significantly more memory-efficient than even mobile models like MobileNet, SqueezeNet, and DenseNet, and approximately as power-efficient as background apps. 
We also include dashed rooflines from other mobile System-on-a-Chip (SoC) and GPUs to compare their peak performance and maximum memory bandwidth. Our model is compatible with most chips, particularly mobile chips such as the Snapdragon 888 and Kirin 9000.

\section{Implementation}

We conducted comprehensive pre-training and a user study as evaluation, covering various scenarios of 14 users, 4 rooms, 19 gestures, different outfits, and remounting. 
This section details the implementation per motion protocol, data collection procedures, hardware/software setup, and evaluation metric for benchmarking \oursys{} against state-of-the-art (SOTA) IOBT utilizing various modalities.

\textbf{Motion protocol:}
\begin{figure}
    \centering
    \includegraphics[width=0.55\linewidth]{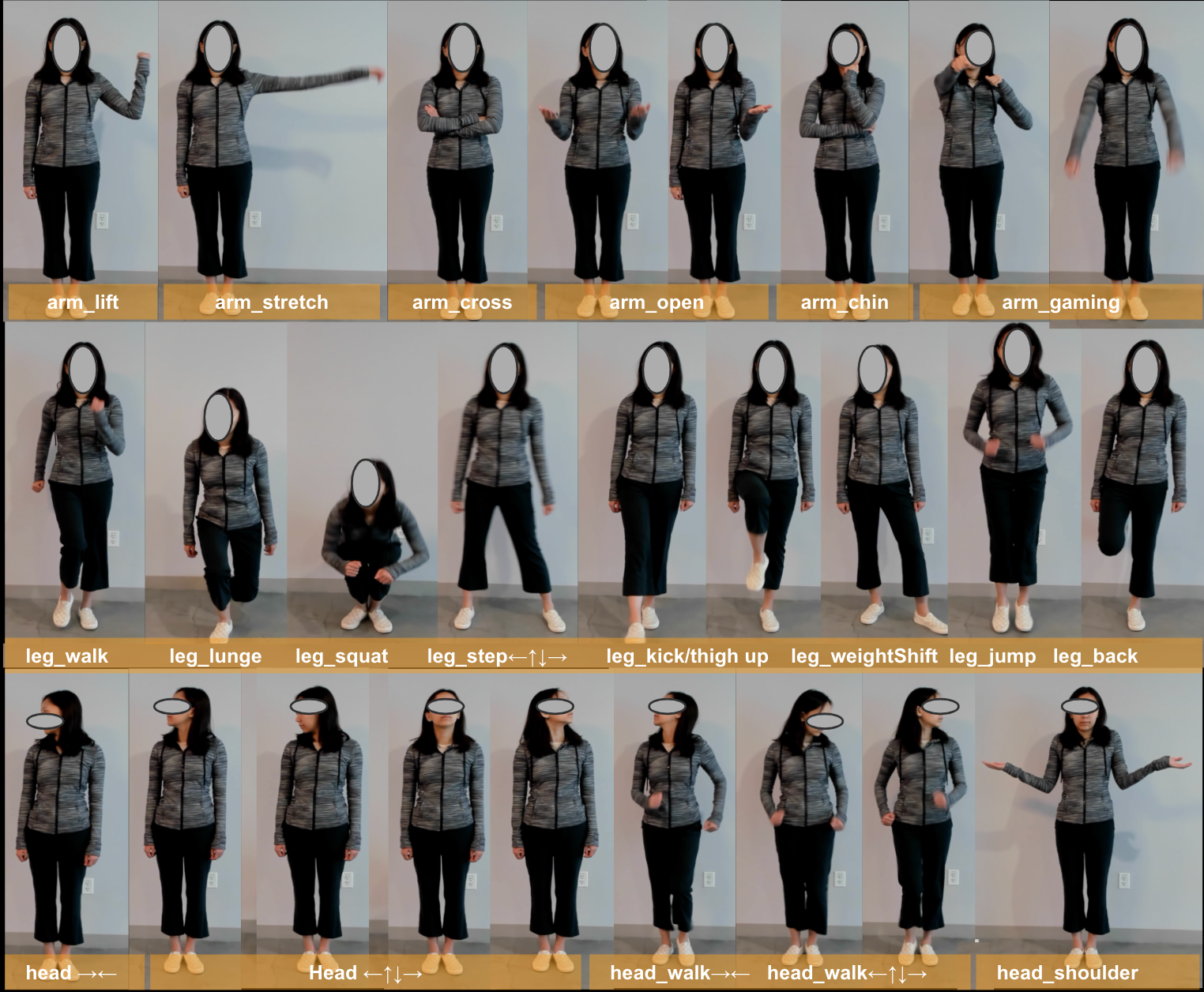}
    \vspace{-7pt}
    \caption{Motion protocol of 19 gestures.}
    \vspace{-10pt}
    \label{fig:motionProtocol}
\end{figure}
Fig.~\ref{fig:motionProtocol} shows 19 selected motions, with emphasis on non-verbal conversational gestures (humans are blocked per the double-blind requirement). To highlight our advantages over cameras, we include challenging occluded gestures like hand-body contact and occluded leg movements.
To ensure a fair comparison, our motion protocol is primarily similar to EgoBody3M~\cite{zhao2025egobody3m}, the first large-scale real-image dataset designed for IOBT on a realistic VR headset. We extend it with additional camera-challenging leg motions and incorporate gestures from mmEgo~\cite{mmEgo}, nearly doubling the number of gestures.
Motions are categorized into arm, leg, and head. Each motion is repeated with the left/right limb at fast/slow speeds, with users naturally recovering to standing pose between repetitions. They are:
1) arm\_lift: lift one arm.
2) arm\_stretch: stretch one arm far.
3) arm\_cross: cross arms for around 1s or 5s.
4) arm\_chin: stroke chin with one hand.
5) arm\_open: open arms with hands up or in a triangle; single or both hands.
6) arm\_gaming: play Beat Saber by chopping/punching with single or both arms.
7) leg\_walking: walk in place with arms swinging in front of chest.
8) leg\_lunge: lunge with one leg.
9) leg\_squat: squat and stand.
10) leg\_step: one step forward, backward, left, or right.
11) leg\_kick: kick one foot forward, then raise one thigh up.
12) leg\_weightShift: plant feet, shift body weight left/right with and without recovering to standing in between.
13) leg\_jump: jump in place with arms in free form.
14) leg\_back: kick one foot back.
15) head\_staticLR: look left or right.
16) head\_static: look left-up, left-down, right-up, right-down, center-up, center-down.
17) head\_walkingLR: look left or right plus walking in place.
18) head\_walking: head\_static plus walking in place.
19) head\_IDK: raise shoulders and open hands, recover, then repeat with head shake.

\textbf{Data collection design - pre-training and user study:}
\label{section: dataCollectionDesign}
To align with real-world use cases, data collection has two stages: pre-training and user study.
1) The pre-training corresponds to a product's development phase, where developers gather an extensive dataset from volunteers over extended periods to build a pre-trained model.
2) User study simulates deployment, where end-users use a product with no training data or minimal calibration/adaptation data.
We recruited 4 participants for pre-training and 10 for user study, under IRB approval. Totally, we collected approximately 525,000 frames of radar data paired with image ground truth, which is 292 minutes at 30 FPS. 
In pre-training, each of the 4 users completed 38 one-minute sessions (2 per gesture.
In user study, each of the 10 participants conducted 2 ten-minute sessions for calibration and testing separately. In each session, they follow a guiding video covering all 19 gestures. Besides these 10 test users, the 4 pre-training participants also did the user study for benchmarking.


\textbf{Experiment setup - hardware:}
User wears the device while standing on the ground, with a camera and a laptop placed 3 meters away. The camera records ground truth exocentric video. The laptop plays a guiding video of motions for the user to follow. Both the headset and camera are USB-connected to a laptop for synchronized data collection. Sensor's details are in the above section. This setup enables synchronous ground truth data that is crucial for training. The guiding video ensures consistent motions execution, facilitating the comparison of results across users and sessions.

\textbf{Experiment setup - software:}
Radars have dedicated MCUs, USB-linked to a laptop. In the future, we aim to integrate these components into XR headsets for wireless, on-device operation. Camera connecting to the same laptop ensures synchronization via a shared system clock. Timestamps from radar and camera are both logged for precise alignment.
Another laptop plays guiding video, triggered simultaneously via a Flask RESTful API. Training and analysis are done offline, using radar SDK v2.4~\cite{infineonRadar}.

\textbf{Evaluation metrics:}
Our evaluation metric for IOBT is the Mean Per-Joint Position Error (MPJPE), a standard metric across sensing modalities. It is calculated as:
$\frac{1}{N} \sum_{i=1}^{N} \left( \frac{1}{M} \sum_{j=1}^{M} \left\lVert \mathbf{p}_{i,j} - \hat{\mathbf{p}}_{i,j} \right\rVert_2 \right)$
where $N$ is the number of frames, $M$ is the number of joints, $P_{i,j}$ is the ground-truth position of the $i-th$ frame and the $j-th$ joint, and $\hat{P}_{i,j}$ is the prediction.
While MediaPipe normalizes it by body height, we rescale using the average participant height (from an optional post-study survey) for fair comparison with SOTA.

\section{Experiment results}

Following the pre-training and user study design, we evaluate across users and vary the amount of user training data. This includes user-independent, user-adaptive, and user-dependent settings, as summarized in Table~\ref{tab:exp-user-3cases}.
\textbf{1) User-independent} means testing on a new user who has no data in training, representing a real-world deployment without prior calibration - the most challenging scenario. We adopt a leave-one-user-out approach, using the other test users for fine-tuning. The average MPJPE is 9.85 cm.
\textbf{2) User-adaptive} allows a minimal calibration/fine-tuning for a new user with their data. It is a more practical setting, aiming to enhance and personalize the device's performance. The MPJPE improves significantly to 4.94 cm.
\textbf{3) User-dependent} requires a large amount of training data from the testing user, showing the upper-bound performance when ample personalization data is available, although not empirical.
We test with pre-training users' data by collecting their user-study protocol. The result is 3.86cm, but testing on a different day with remounting and outfit changes increases the error to 7.21 cm, revealing sensitivity to such variations.


To contextualize our results with the SOTA, the exocentric 3D human pose estimation, which we rely on as ground truth, reports top-5 performance ranging from 17.59mm to 29.1mm MPJPE on the Human3.6M benchmark~\cite{chun2023learnable, he2020epipolar}. EgoBody3M~\cite{zhao2025egobody3m} uses an egocentric camera with an MPJPE of 5.18cm, but a significantly higher 12.2cm error for out-of-view wrists. The mmWave+IMU egocentric work~\cite{mmEgo} reports 4.287cm for upper body and 5.460cm for lower body, though it was tested on only 3 users. In comparison, our user-adaptive result (4.94cm) demonstrates performance comparable to SOTA egocentric/IOBT approaches. Moreover, we train on a dataset orders of magnitude smaller and offer advantages such as reduced error per body occlusions and a more compact hardware design for wearables.

\begin{table}[H]
    \centering
    \vspace{-7pt}
    \begin{tabular}{|l|l|}
    \hline
    \textbf{} & MPJPE/cm \\ \hline
    User-independent & 9.85 \\ \hline
    User-adaptive & 4.94 \\ \hline
    User-dependent & 3.86 (7.21 remount\&cross-outfit) \\ \hline
    \end{tabular}
    \vspace{-7pt}
    \caption{Results for testing user generalizability.}
    \vspace{-20pt}
    \label{tab:exp-user-3cases}
\end{table}

\begin{figure*}
    \centering
    \begin{subfigure}[t]{0.32\linewidth}
        \centering
        \includegraphics[width=0.95\textwidth]{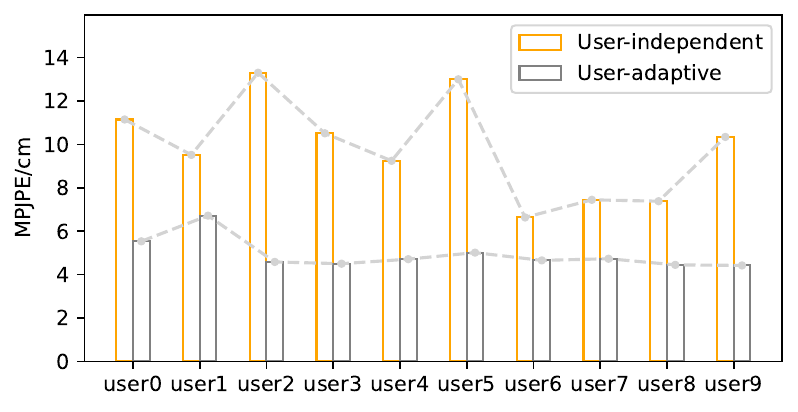}
        \vspace{-7pt}
        \caption{User-independent V.S. User-adaptive.}
        \label{fig:exp_protocol1}
    \end{subfigure}
    \hfill
    \begin{subfigure}[t]{0.35\linewidth}
        \centering
        \includegraphics[width=0.88\textwidth]{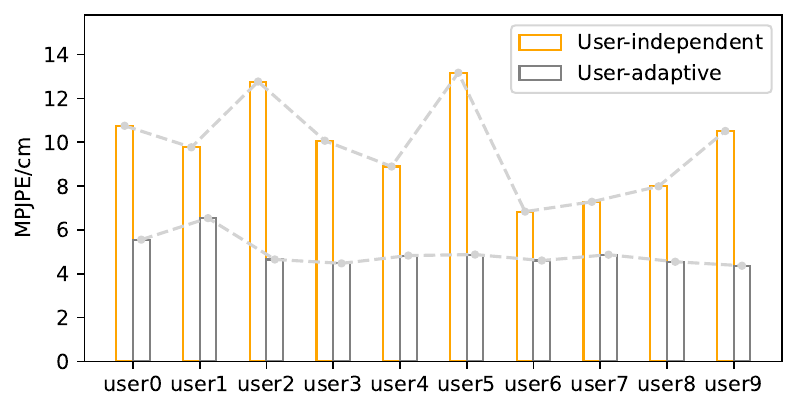}
        \vspace{-7pt}
        \caption{User-independent V.S. adaptive (2 protocols).}
        \label{fig:exp_protocol2}
    \end{subfigure}
    \hfill
    \begin{subfigure}[t]{.32\linewidth}
        \centering
        \includegraphics[width=0.98\textwidth]{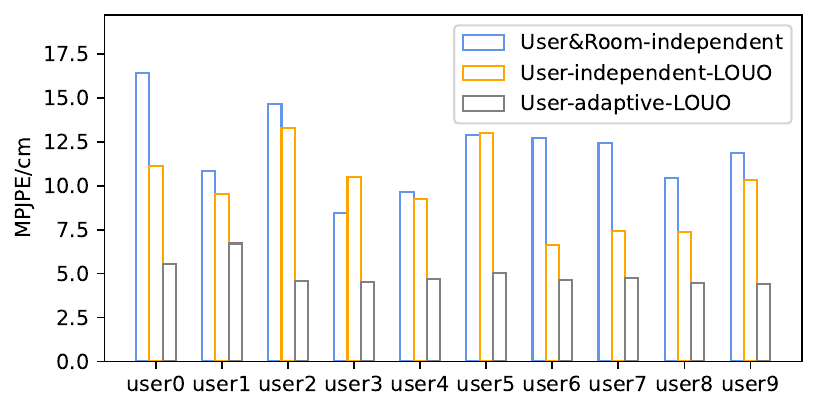}    
        \vspace{-7pt}
        \caption{User\&Room-independent.}
        \label{fig:exp_room-user-wise}
    \end{subfigure}

    \vspace{5pt}
    \begin{subfigure}[t]{0.19\linewidth}
        \centering
        \includegraphics[width=0.9\textwidth]{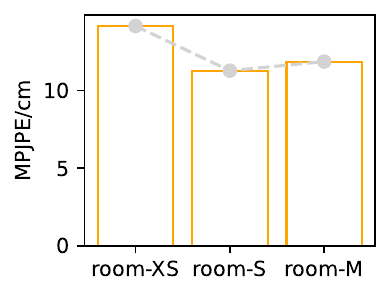}
        \vspace{-7pt}
        \caption{User\&Room-indep-\\endent(room-wise).}
        \label{fig:exp_room}
    \end{subfigure}
    \hfill
    \begin{subfigure}[t]{0.215\linewidth}
        \centering
        \includegraphics[width=.97\textwidth]{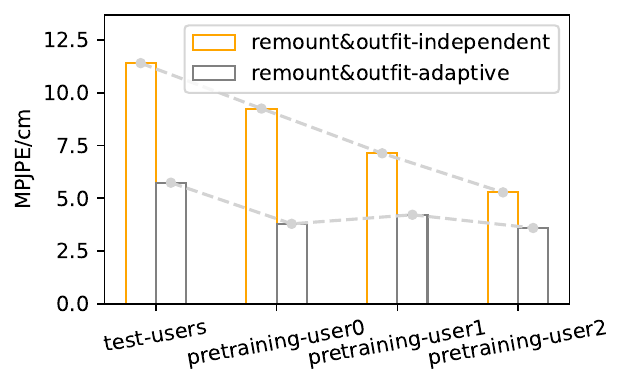}
        \vspace{-7pt}
        \caption{Remount\&outfit change.}
        \label{fig:exp_remountOutfit}
    \end{subfigure}
    \hfill
    \begin{subfigure}[t]{0.29\linewidth}
        \centering
        \includegraphics[width=.97\textwidth]{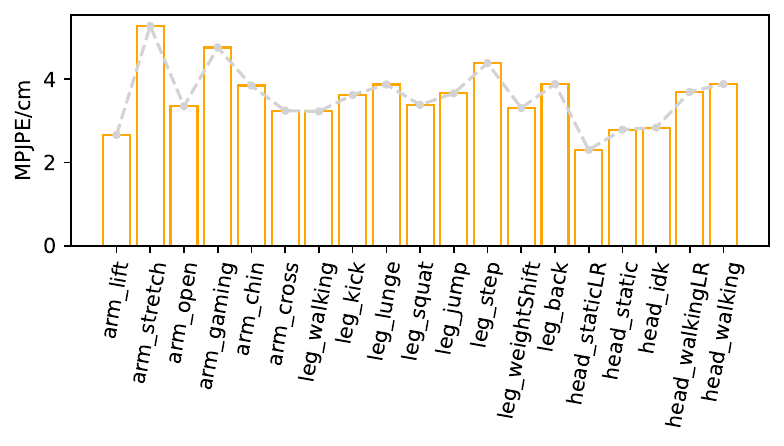}
        \vspace{-7pt}
        \caption{Gesture-wise error analysis.}
        \label{fig:exp_gesture}
    \end{subfigure}
    \hfill
    \begin{subfigure}[t]{0.29\linewidth}
        \centering
        \includegraphics[width=.97\textwidth]{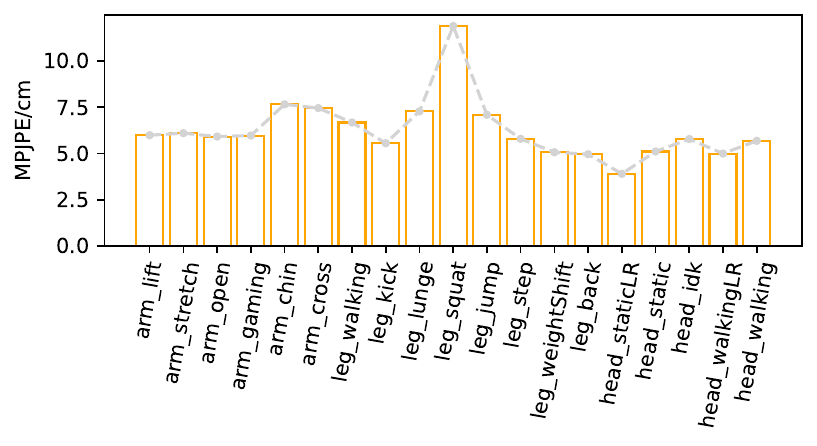}
        \vspace{-7pt}
        \caption{Leave-one-gesture-out.}
        \label{fig:exp_LOGO}
    \end{subfigure}

    \vspace{-7pt}
    \caption{Error analysis and micro-benchmarking from different perspectives.}
    \label{fig:breakdown}
    \vspace{-15pt}
\end{figure*}

\subsection{Error analysis and micro-benchmarking}
\textbf{}

We also break down the errors for analysis and evaluate with micro-benchmarking experiments to show that \oursys is robust across gestures, environment/rooms, remounting and outfit, and user adaptation.

\textbf{Gesture generalizability:}
1) Gesture-wise error analysis:
(using only the user-adaptive results to isolate irrelevant factors.)
As in Fig.~\ref{fig:exp_gesture}, arm\_stretch gesture is the worst, possibly due to the limited FoV of ground truth camera. Additionally, arm\_gaming may be challenging due to its higher speed.
Moreover, there is an upward trend in error for head gestures since they are complex. 
Most other gestures perform well, suggesting our capability to track diverse gestures, even camera-challenging leg motions and hand-body contacts.
2) Effect of unseen gesture:
Performance on an unseen gesture can be tested in a leave-one-gesture-out manner in pre-training.
In Fig.~\ref{fig:exp_LOGO}, leg\_squat yields a high error, possibly due to its large motion range. The overall results demonstrate a reasonably small performance drop, indicating strong generalizability to unseen gestures as long as training covers diverse motions as we do.

\begin{figure}[H]
    \vspace{-10pt}
    \centering
    \includegraphics[width=0.65\linewidth]{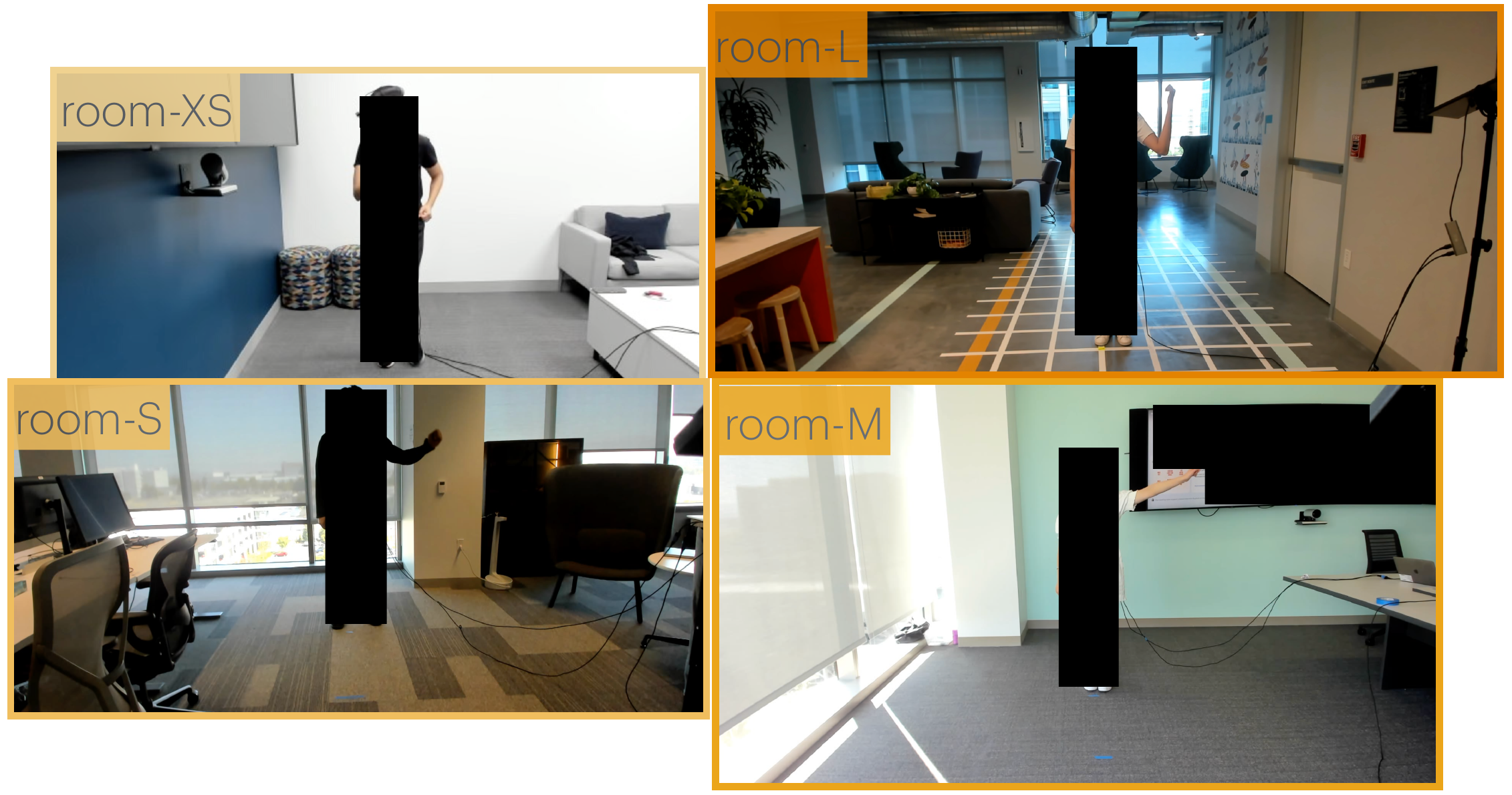}
    \vspace{-10pt}
    \caption{Various data collection rooms.}
    \label{fig:exp-rooms-photo}
    \vspace{-10pt}
\end{figure}

\textbf{Environment generalizability: }
We collected data in 4 different rooms, as illustrated in Fig.~\ref{fig:exp-rooms-photo}. (Humans are blocked.)
Pre-training is in a large room (L). User study is in medium (M), small (S), and extra-small (XS) rooms.
They are typical office or home settings, with furniture located 1-2 meters away from users. 
To isolate the effect of rooms, we evaluate a user\&room independent test by excluding the same room when fine-tuning in a leave-one-user-out(LOUO) manner. As shown in Fig.~\ref{fig:exp_room-user-wise}, room adaptation improves performance, though less than user adaptation.
A room-wise error analysis (Fig.~\ref{fig:exp_room}) shows minimal differences across room sizes, demonstrating our robustness and suitability for deployment in diverse indoor environments.

\textbf{Remounting and outfit: }
To assess these effects, we had a small additional data collection from the same user on another day, involving pre-training participants using the user study protocol. This user-dependent study isolates the impact of remounting and outfit change from other factors.
Fig.~\ref{fig:exp_remountOutfit} shows the user-wise results, alongside test participants' average user-independent/adaptive results; this presents a comparison between with and without adaptation. Notably, both show a similar accuracy degradation.
This indicates that the cross-user degradation is highly likely an inevitable consequence of remounting and outfit.

\textbf{Effectiveness of adaptation: }
To evaluate the effectiveness of user adaptation, we conducted a user-wise analysis. Fig.~\ref{fig:exp_protocol1} shows that adaptation consistently improved accuracy across all users.
Furthermore, we fine-tune with a small amount of user-study data from pre-training users. Fig.~\ref{fig:exp_protocol2} illustrates that this does not lead to noticeable improvements for testing users.
So, the benefits of user adaptation stem from adapting to individual-specific characteristics, rather than from data volume or recipes of data sequence. This highlights the importance of user adaptation in achieving high accuracy in gesture recognition systems.


\subsection{Visualization}

\begin{figure*}[!ht]
    \begin{minipage}{0.67\textwidth}
        \centering
        \includegraphics[width=.73\linewidth]{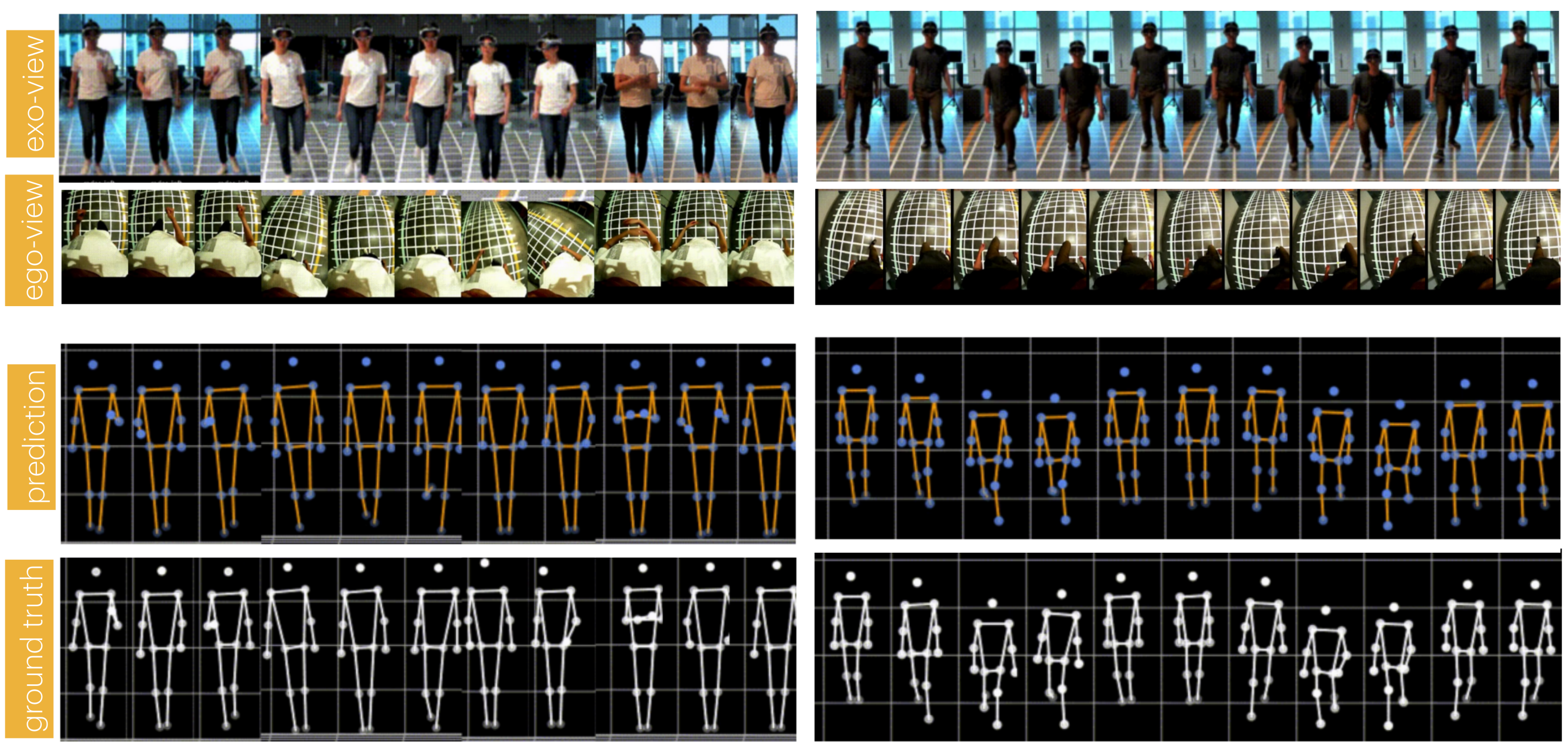}
        \vspace{-5pt}
        \caption{Visualizations of our results alongside exo-/ego-camera views. (Exo-view is our ground truth camera; Ego-view is only used for comparison.)}
        \vspace{-5pt}
        \label{fig:visualizations}
    \end{minipage}
    \hfill
    \begin{minipage}{0.3\textwidth}
        \vspace{20pt}
       \centering
        \includegraphics[width=\linewidth]{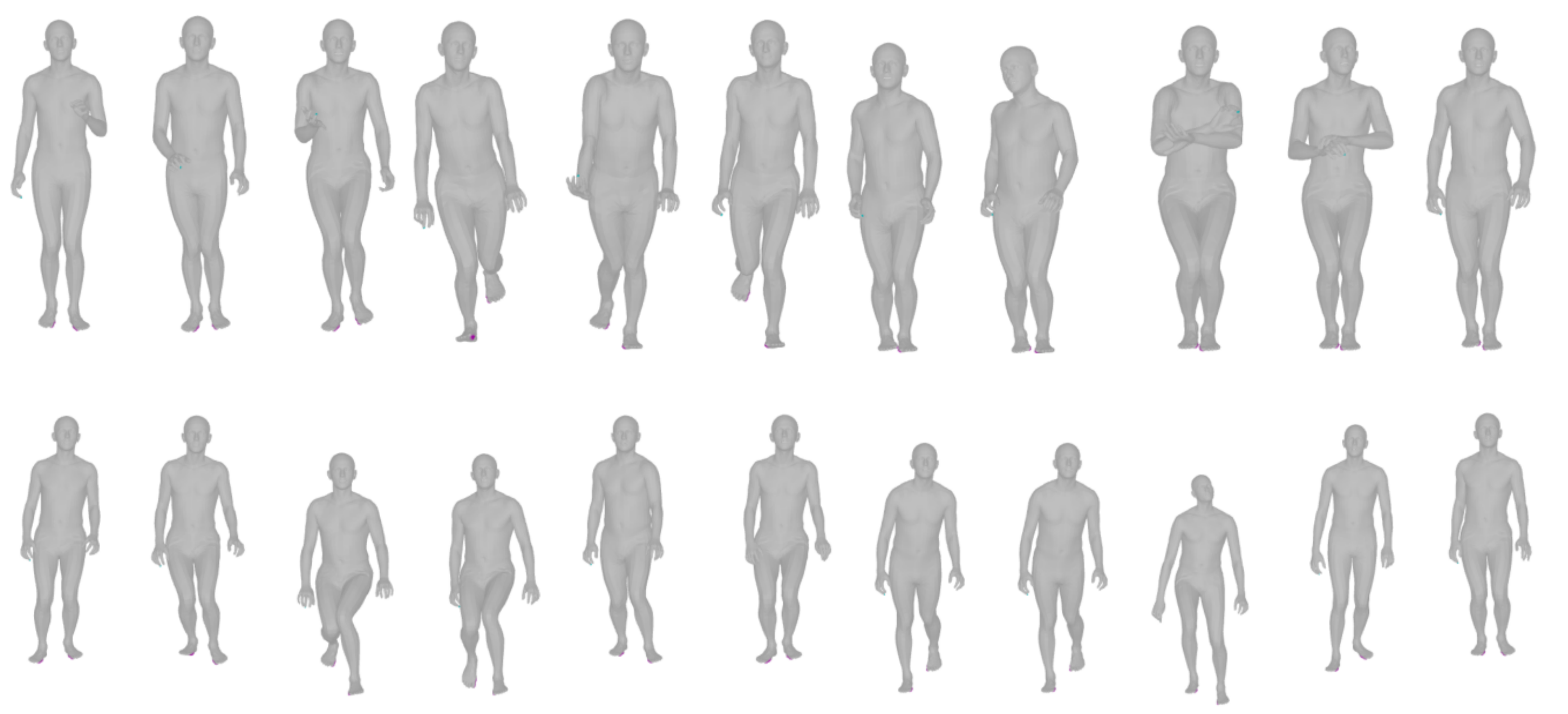}
        \vspace{-5pt}
        \caption{Output keypoints as an encoder can drive a downstream task of codec avatar.}
        \vspace{-5pt}
        \label{fig:smpl}
    \end{minipage}
    \vspace{-7pt}
\end{figure*}

To interpret our results, we provide visualizations of predictions alongside the ground truth and sample camera views (both egocentric and exocentric) in Fig.~\ref{fig:visualizations}. Screenshots showcase representative gestures including walking in place, head rotation, crossing arms, and squatting.
We particularly include a plot of a challenging leg motion at a higher frame rate for the right user. Specifically, the egocentric view of kicking back is fully obscured due to self-occlusion issues posed by egocentric cameras. However, our radar can see through body occlusion and make accurate predictions in camera-challenging cases.
In this case, the egocentric view is fully obscured due to self-occlusion, a common issue for cameras, yet our radar reliably captures the motion.
Besides, a video in the supplementary material provides a more detailed, dynamic view of these results.

\subsection{Demo downstream task: Body Codec Avatar}

IOBT supports key tasks in MR, such as gesture control and driving codec avatars. Recently, codec avatar becomes particularly important for natural, photo-realistic presence and interaction in MR. 
Here, IOBT can serve as the encoder (transmitter code) for body tracking to drive a codec avatar in real-time, which combines with environmental lighting and receiver view.
To achieve this, our predicted body keypoints first map to a skeleton, then refined by inverse kinematics (IK) as model parameters. 
These parameters can then render a codec avatar using tools like Codec Avatar Studio or common SMPL-related libraries.
Since we lack a full 3D scan of participants as the decoder (typically obtained from a camera dome), we instead use a SMPL-X 3D mesh model without texture--a widely used body representation. Fig.~\ref{fig:smpl} is a demonstration based on the selected frames in Fig.~\ref{fig:visualizations}.
In future work, we could tailor a self-supervised encoder learning pipeline to fuse this new radar modality end-to-end.
\section{Related work}

\textbf{Wireless perception:} Recent advances in wireless perception using RF or ultrasound signals have enabled various applications. 
Google's Soli ~\cite{hayashi2021radarnet, wang2016interacting} uses a custom miniature mmWave radar for gesture classification on mobile phones. WiSee ~\cite{pu2013whole} pioneered home-scale gesture recognition system using WiFi, followed by~\cite{zhang2021widar3, qian2018widar2, qian2017widar}. These systems support only limited pre-defined gestures. 
Beyond this, fine-grained wireless perception has been explored, such as face reconstruction~\cite{xie2023mm3dface} using mmWave, motion detection and exocentric pose estimation using expensive software-defined radios like USRP~\cite{adib2013see, adib20143d, adib2015rf, zhao2018rf, zhao2018through}, and 3D exocentric pose estimation using commercial WiFi~\cite{jiang2020towards}.

\textbf{mmWave body tracking: }
Recent mmWave human estimation work focuses on the exocentric~\cite{singh2019radhar, zhao2023cubelearn} and datasets~\cite{marsdataset2021, an2022mri, cui2024milipoint, yang2024mm}, or synthetic data using generative models~\cite{zhao2023nerf2, chen2023rfgenesis, 2023synthesizemili}. However, the egocentric/IOBT is much less explored even in computer vision field.
mmWave's advantage in seeing through NLOS is particular critical in egocentric body tracking/IOBT, where occlusion is much worse than exocentric perception. 
Recently,~\cite{mmEgo} pioneered IOBT using head-mounted mmWave radars and IMU sensors. Their prototype is a helmet with radars positioned around 40cm away from the face. While far positioning broaden sensors' body view, it compromises ergonomics and form factor, rendering the system impractical for commercial MR devices. Their error is 4.287cm for upper body and 5.460cm for lower body, but limited to only three users' data. \cite{duan2025argus} placed radars above the shoulders, around 15cm to the ear on a headphone, and their approach was also empirical.
In contrast, \oursys{} is the first system to mmWave IOBT specifically designed for MR headsets, which is only 4cm away from cheek and requires no additional sensors like IMU.

\textbf{Camera-based egocentric body tracking.} 
Unlike exocentric body tracking, IOBT has only recently gained attention.
EgoCap~\cite{rhodin2016egocap} was among the first to propose camera-based IOBT using two cameras mounted on rigid rods attached to a bicycle helmet. EgoGlass~\cite{zhao2021egoglass} suggests a more practical setup, with two cameras located on the temples of a pair of glasses.
Besides, the scarcity and difficulty of collecting egocentric data have led to the wide use of synthetic data like Mo$^2$Cap$^2$~\cite{xu2019mo2cap2} and EgoPose~\cite{tome2019xr}. Despite that their camera placements—around 8cm from the head or beneath a VR headset—are impractical.
EgoBody3M \cite{zhao2025egobody3m} is a new dataset captured with Meta's MR headset with a combination of outside-in views in the wild. We adopt their approach of using exocentric/outside-in cameras to obtain accurate 3D ground truth.
However, Meta Orion~\cite{orion2024} envisions compact glasses for MR; in the meantime, pancake lenses~\cite{zhao2025egobody3m} used in Meta's Quest 3 and Apple's Vision Pro, reduce headset thickness, making it more challenging for future development of camera-based IOBT system. 
Recent research starts seeking opportunities by leveraging powerful generative models to infer full body pose using only sparse input from ego-view.
In contrast, radar offers a compact, occlusion-robust modality, meeting the needs for future IOBT for MR.

\section{Limitation and future work}
Usually, a separate encoder serves keypoint coordinates as a skeleton parameter to drive a 3D model, which can lead to unrealistic appearances of the 3D model, such as twisting joints. Recent trends in 3D computer vision aim to develop more end-to-end encoder and decoder models for codec avatars, eliminating intermediate parameters such as joint angles and pose parameters, and instead learning a direct mapping from sensor data to 3D mesh model parameters. Or we could propose new methods to tailor an unsupervised encoder learning pipeline to radar tracking without pixel-by-pixel input images to compute loss, enabling an end-to-end system fusing this new encoder modality.

While this paper demonstrates the feasibility of mmWave as a modality for IOBT, sensor fusion with cameras might be a more practical future direction. This is because cameras can provide high-resolution visual information that can complement the radar data, especially in scenarios where the radar signal is weak or occluded. For example, in situations where the user is sitting on a chair with their legs under the table, the radar signal may be blocked, but a camera could still capture the user's leg movements.

Radar has the potential to work on some extreme corner cases that cameras will completely fail, such as big belly fully occluding the legs, or wearing a large coat that occlude the lower body. In these cases, the radar signal can penetrate through obstacles and provide accurate body tracking, whereas cameras would be unable to capture any meaningful data. Future work could explore the use of radar in these extreme scenarios and develop methods to robustly track the body even in the presence of significant occlusions.



\bigskip

\bibliography{main}
\end{document}